\documentclass[12pt,thmsa]{article}
\usepackage{amsmath}
\usepackage{amsfonts}

\input{tcilatex}
\begin{document}

\author{Diego Julio Cirilo-Lombardo \\
{\small Bogoliubov Laboratory of Theoretical Physics}\\
{\small Joint Institute for Nuclear Research, 141980, Dubna, Russian
Federation.}\\
{\small diego@thsun1.jinr.ru ; diego77jcl@yahoo.com}}
\title{On the mathematical structure and hidden symmetries of the
Born-Infeld field equations}
\date{The Date }
\maketitle

\begin{abstract}
The mathematical structure of the Born-Infeld field equations was analyzed
from the point of view of the symmetries. To this end, the field equations
were written in the most compact form by means of quaternionic operators
constructed according to all the symmetries of the theory, including the
extension to a non-commutative structure. The quaternionic structure of the
phase space was explicitly derived and described from the Hamiltonian point
of view, and the analogy between the BI\ theory and the Maxwell (linear)
electrodynamics in curved space-time was explicitly shown. Our results agree
with the observation of Gibbons and Rasheed that there exists a discrete
symmetry in the structure of the field equations that is unique in the case
of the Born-Infeld nonlinear electrodynamics.
\end{abstract}

\section{Introduction}

\bigskip The most significant nonlinear theory of electrodynamics is mainly the Born-Infeld (BI) theory. The Lagrangian density describing
the BI theory (in arbitrary space-time dimensions) is
\begin{equation}
\mathcal{L}_{BI}=\sqrt{-g}L_{BI}=\frac{b^{2}}{4\pi }\left\{ \sqrt{-g}-\sqrt{%
-\det (g_{\mu \nu }+b^{-1}F_{\mu \nu })}\right\}  \tag{1}
\end{equation}%
where $b$ is a fundamental parameter of the theory with field dimensions%
\footnote{%
In open superstring theory (2-dimensions), for example, loop calculations
lead to this Lagrangian with $b^{-1}=2\pi \alpha ^{\prime }$ ($\alpha
^{\prime }\equiv \ $inverse of the string tension)}. In four space-time
dimensions the determinant in (1) may be expanded out to give
\begin{equation}
L_{BI}=\frac{b^{2}}{4\pi }\left\{ 1-\sqrt{1+\frac{1}{2}b^{-2}F_{\mu \nu
}F^{\mu \nu }-\frac{1}{16}b^{-4}\left( F_{\mu \nu }\widetilde{F}^{\mu \nu
}\right) ^{2}}\right\}  \tag{2}
\end{equation}%
which coincides with the usual Maxwell Lagrangian in the weak field limit.
Similarly, if we consider the second rank tensor $\mathbb{F}^{\mu \nu }$
defined by
\begin{equation}
\mathbb{F}^{\mu \nu }=-\frac{1}{2}\frac{\partial L_{BI}}{\partial F_{\mu \nu
}}=\frac{F^{\mu \nu }-\frac{1}{4}b^{-2}\left( F_{\rho \sigma }\widetilde{F}%
^{\rho \sigma }\right) \,\widetilde{F}^{\mu \nu }}{\sqrt{1+\frac{1}{2}%
b^{-2}F_{\rho \sigma }F^{\rho \sigma }-\frac{1}{16}b^{-4}\left( F_{\rho
\sigma }\widetilde{F}^{\rho \sigma }\right) ^{2}}}  \tag{3}
\end{equation}%
(so that $\mathbb{F}^{\mu \nu }\approx F^{\mu \nu }$ for weak fields), this
second kind of antisymmetrical tensor satisfies the electromagnetic
equations of motion
\begin{equation}
\nabla _{\mu }\mathbb{F}^{\mu \nu }=0  \tag{4}
\end{equation}%
which are highly nonlinear in $F_{\mu \nu }$. Another interesting object to
analyze is the energy-momentum tensor that can be written as
\begin{equation}
T_{\mu \nu }=\frac{1}{4\pi }\left\{ \frac{F_{\mu }^{\,\ \ \lambda }F_{\nu
\lambda }+b^{2}\left[ \mathbb{R}-1-\frac{1}{2}b^{-2}F_{\rho \sigma }F^{\rho
\sigma }\right] g_{\mu \nu }}{\mathbb{R}}\right\}  \tag{5}
\end{equation}%
\begin{equation*}
\mathbb{R}\equiv \sqrt{1+\frac{1}{2}b^{-2}F_{\rho \sigma }F^{\rho \sigma }-%
\frac{1}{16}b^{-4}\left( F_{\rho \sigma }\widetilde{F}^{\rho \sigma }\right)
^{2}}
\end{equation*}%
However, besides these more or less obvious statements, Born and Infeld
observed in their original work [1] that the tensor $\mathbb{F}$ has a relation
to $F$ similar to that in Maxwell theory of macroscopical bodies
between the dielectric displacement and magnetic induction, but in the BI case
this relation is a discrete electric-magnetic duality invariance[5] that is
associated with underlying $SO\left( 2\right) $ symmetry. In ref.[1], the
relations that put in evidence the symmetries of these transformations that
are characteristic of the BI\ field equations only, are%
\begin{equation}
\mathbb{F}^{\mu \nu }=\frac{F^{\mu \nu }-G\widetilde{F}^{\mu \nu }}{\sqrt{%
1+S-G^{2}}}  \tag{6}
\end{equation}%
\begin{equation}
\ \ \ \ \ \ \ \ \ \ \ \ \ \ \ F^{\mu \nu }=\frac{\mathbb{F}^{\mu \nu }+Q%
\widetilde{\mathbb{F}}^{\mu \nu }}{\sqrt{1+P-Q^{2}}},\ \ Q\equiv G  \tag{7}
\end{equation}%
where $G,\ Q,\ S$ and $P$ are the electromagnetic invariants constructed of
the two types of fields $F$ and $\mathbb{F}$ which will be expressed
explicitly in the next section. Although it is by no means obvious, it may
be verified that equations (3), (6) and (7) are invariant under
electric-magnetic rotations of duality $F\longleftrightarrow \widetilde{%
\mathbb{F}}$, but notice that the BI\ Lagrangian (1) is not. This fact was
firstly pointed out in the general publications about the electromagnetic
duality rotations of Gaillard and Zumino[4] and more recently and
specifically for the BI case, in the papers of Gibbons and Rasheed [5,6].

The main task in this work is to complete in any sense the analysis given in
refs.[4,5,6] for the BI\ theory showing explicitly the quaternionic
structure of the field equations. The starting point to complete such an
analysis is based on a previous paper of the author [7] where it was
explicitly shown that the transformations (6,7) are produced by a
quaternionic operator acting over vectors in which the components are the
corresponding electromagnetic fields%
\begin{equation}
\left(
\begin{array}{l}
\mathbb{F} \\
\\
\widetilde{\mathbb{F}}%
\end{array}%
\right) ^{\mu \nu }=\overset{\equiv A}{\overbrace{\frac{1}{\mathbb{R}}\left(
\sigma _{0}-i\sigma _{2}\mathbb{P}\right) }}\left(
\begin{array}{l}
F \\
\\
\widetilde{F}%
\end{array}%
\right) ^{\mu \nu }  \tag{8}
\end{equation}

\begin{equation}
\left(
\begin{array}{l}
F \\
\\
\widetilde{F}%
\end{array}%
\right) ^{\mu \nu }=\overset{\equiv \overline{B}}{\overbrace{\frac{\mathbb{R}%
}{1+\mathbb{P}^{2}}\left( \sigma _{0}+i\sigma _{2}\mathbb{P}\right) }}\left(
\begin{array}{l}
\mathbb{F} \\
\\
\widetilde{\mathbb{F}}%
\end{array}%
\right) ^{\mu \nu },  \tag{9}
\end{equation}%
where $\widetilde{\mathbb{F}}^{kl}\equiv \frac{\partial L_{BI}}{\partial
\widetilde{F}_{kl}}$ ; $\mathbb{R}\equiv \sqrt{1+\frac{1}{2}b^{-2}F_{\rho
\sigma }F^{\rho \sigma }-\frac{1}{16}b^{-4}\left( F_{\rho \sigma }\widetilde{%
F}^{\rho \sigma }\right) ^{2}}$ and $\mathbb{P}=-\frac{1}{4b^{2}}F_{\mu \nu }%
\widetilde{F}^{\mu \nu }$ $\left( b\equiv \text{absolute field of the
BI-theory}\right) $ and the complex conjugation indicated by the
horizontal-bar over the operators. The Pauli matrices are defined as
(Landau-Lifshitz 1968)%
\begin{equation*}
\begin{array}{cccc}
\sigma _{1}=\left(
\begin{array}{cc}
0 & 1 \\
1 & 0%
\end{array}%
\right) & \sigma _{2}=\left(
\begin{array}{cc}
0 & -i \\
i & 0%
\end{array}%
\right) & \sigma _{3}=\left(
\begin{array}{cc}
1 & 0 \\
0 & -1%
\end{array}%
\right) & \sigma _{0}=\left(
\begin{array}{cc}
1 & 0 \\
0 & 1%
\end{array}%
\right)%
\end{array}%
\end{equation*}%
The norms of the operators $A$ and $B$ are

\begin{equation*}
\overline{A}A=A\overline{A}=\frac{1+\mathbb{P}^{2}}{\mathbb{R}^{2}}
\end{equation*}

\begin{equation*}
\overline{B}B=B\overline{B}=\frac{\mathbb{R}^{2}}{1+\mathbb{P}^{2}}
\end{equation*}%
where from expressions (8,9) we have

\begin{equation*}
A\overline{B}=\overline{A}B=1
\end{equation*}%
The plan of this paper is as follows: in Section 2, the Quaternionic
structure of the BI\ field equations is manifestly presented and the
mathematical structure is carefully analyzed and extended.\ In Section 3, we
describe the phase space determined by the symmetries of the BI\ field
equations from the Hamiltonian point of view. In Section 4, the
constitutive-like relations of the BI\ theory are studied by comparing them
with the ordinary Maxwell electrodynamics in Riemannian space with an
arbitrary metric and the expression for the Fresnel equation is explicitly
given for the BI case. Finally, the remarks and conclusions are given in
Section 5.

Our convention is as in ref.[2] with signatures of the metric, Riemann and
Einstein tensors (-++); the internal indeces (gauge group) are denoted by $%
a,b,c...$, space-time indeces by Greek letters $\mu ,\nu ,\rho ...$ and the
tetrad indeces by capital Latin letters $A,B,C..$

\section{The quaternionic structure}

Now we will show in an explicit and compact form how the transformations
(6,7) can be realized by means of a quaternionic structure. We start with
the following definitions for the invariants of the electromagnetic field $%
S\equiv \frac{1}{2b^{2}}F_{\rho \sigma }F^{\rho \sigma }$, $G=\frac{1}{2b^{2}%
}F_{\mu \nu }\widetilde{F}^{\mu \nu }$. $\mathbb{R\equiv }\sqrt{1+S-G^{2}},$
and the following signature for the metric tensor is adopted $g_{\mu \nu
}=\left( ---+\right) $ Starting from expressions (6,7)\ with the new
definitions for the invariants we have%
\begin{equation}
\left(
\begin{array}{l}
\mathbb{F} \\
\\
\widetilde{\mathbb{F}}%
\end{array}%
\right) ^{\mu \nu }=\overset{\equiv \mathbb{Q}}{\frac{1}{\mathbb{R}}%
\overbrace{\left( \sigma _{0}-i\sigma _{2}G\right) }}\left(
\begin{array}{l}
F \\
\\
\widetilde{F}%
\end{array}%
\right) ^{\mu \nu }  \tag{10}
\end{equation}%
It is interesting to notice that, due to the fact that the following
identity holds $F_{\mu \nu }\widetilde{F}^{\mu \nu }=\mathbb{F}_{\mu \nu }%
\widetilde{\mathbb{F}}^{\mu \nu },$ the quaternion $\mathbb{Q}$ is \textit{%
invariant }from the topological point of view. It is a very important
property because the mapping between the different sets of fields, $F$ and $%
\mathbb{F}$, respectively, preserves the topological charge unaltered. This
means that the topological charge is a fixed point of the $\mathbb{Q}$
transformation. Defining the \textquotedblleft spinors\textquotedblright\
\begin{equation*}
\begin{array}{cc}
\Psi =\left(
\begin{array}{l}
F \\
\\
\widetilde{F}%
\end{array}%
\right) & \overline{\Psi }=\left( \sigma _{3}\Psi \right) ^{\dag }%
\end{array}%
,
\end{equation*}%
and
\begin{equation*}
\begin{array}{cc}
\Phi =\left(
\begin{array}{l}
\mathbb{F} \\
\\
\widetilde{\mathbb{F}}%
\end{array}%
\right) & \overline{\Phi }=\left( \sigma _{3}\Phi \right) ^{\dag }%
\end{array}%
,
\end{equation*}%
the square root $\mathbb{R}$ in (10) is simplified to the following
expression:%
\begin{equation*}
\sqrt{1+S-G^{2}}=\sqrt{1+\frac{1}{4}\left( \overline{\Psi }\mathbb{Q}\Psi
\right) }
\end{equation*}%
and relation (10) takes the compact form%
\begin{equation}
\Phi =\frac{\mathbb{Q}\Psi }{\sqrt{1+\frac{1}{4}\left( \overline{\Psi }%
\mathbb{Q}\Psi \right) }}  \tag{11}
\end{equation}%
As we could see in the introduction [1], in the same manner it is possible
to invert the above equation putting all as a function of the spinor $\Psi $%
. In order to do this, it is sufficient to consider: $P\equiv \frac{1}{2b^{2}%
}\widetilde{\mathbb{F}}_{\rho \sigma }\widetilde{\mathbb{F}}^{\rho \sigma },$
$Q=G=\frac{1}{2b^{2}}\mathbb{F}_{\mu \nu }\widetilde{\mathbb{F}}^{\mu \nu }$%
and the following property $F_{\rho \sigma }F^{\rho \sigma }=-\widetilde{F}%
_{\rho \sigma }\widetilde{F}^{\rho \sigma }$. The square root in this
inverted transformation is ($\overline{\mathbb{Q}}\equiv \left( \sigma
_{0}+i\sigma _{2}G\right) $)%
\begin{equation*}
\sqrt{1+P-Q^{2}}=\sqrt{1-\frac{1}{4}\left( \overline{\Phi }\overline{\mathbb{%
Q}}\Phi \right) }
\end{equation*}%
and the inverse transformation becomes%
\begin{equation}
\Psi =\frac{\overline{\mathbb{Q}}\Phi }{\sqrt{1-\frac{1}{4}\left( \overline{%
\Phi }\overline{\mathbb{Q}}\Phi \right) }}  \tag{12}
\end{equation}

We stop here to see in a little more detail the mathematical structure of
the operators $\mathbb{Q}$. From (10) we can see that the $\mathbb{Q}$ form
part of a commutative ring of complex operators $\mathbb{Q}\equiv \left\{
\alpha +i\beta \mathbb{I}\ /\alpha ,\beta \in \mathbb{C}\right\} $ ,
equipped with addition and multiplication laws induced by those in $\mathbb{C%
}$, such as addition and multiplication on $\mathbb{Q}$ are given by the
usual matrix addition \ and multiplication,with $\mathbb{I}\ $having the
following form\footnote{%
here $d$ is the dimension.}:%
\begin{equation*}
\mathbb{I}\ =\pm \left(
\begin{array}{cc}
0 & 1_{d/2} \\
-1_{d/2} & 0%
\end{array}%
\right)
\end{equation*}%
It is easily seen that $\mathbb{Q}$ is a commutative ring with zero divisors%
\begin{equation*}
\mathbb{Q}_{\pm }^{0}\equiv \left\{ \lambda \left( 1_{d}\pm i\mathbb{I}%
_{d}\right) ,\lambda \in \mathbb{C}\right\} ,
\end{equation*}%
$\mathbb{Q}_{-}^{0},\mathbb{Q}_{+}^{0}$ are the only multiplicative ideals
in $\mathbb{Q}$, for instance, are maximal ideals. Thus, the only fields
that we can construct from $\mathbb{Q}$ are%
\begin{equation*}
\frac{\mathbb{Q}}{\mathbb{Q}_{\pm }^{0}}\cong \mathbb{Q}_{\mp }^{0}\cong
\mathbb{C}
\end{equation*}%
In the general case (e.g: $\alpha ,\beta \in \mathbb{C)}$, the map $%
\left\vert \ \cdot \ \right\vert ^{2}:\mathbb{Q}\rightarrow \mathbb{R}%
/\left\vert \mathbb{Q}\right\vert ^{2}\equiv \mathbb{Q}\overline{\mathbb{Q}}%
=\alpha ^{2}+\beta ^{2}$ can be seen as a semi-modulus on the ring $\mathbb{Q%
}$%
\begin{equation*}
\mathbb{Q}=\mathbb{Q}^{+}\cup \mathbb{Q}^{0}\cup \mathbb{Q}^{-},
\end{equation*}%
according to the sign of the modulus of\ $\mathbb{Q}$. It is important to
note that in contrast with the analysis of ref.[9], for the BI case $\alpha
,\beta $ $\in \mathbb{R}$ ($\alpha ,\beta ,$ are the identity and the
pseudoscalar invariant of the electromagnetic field respectively) the
commutative ring described by $\mathbb{Q}\,\ $has no pseudo-complex
structure.

Another interesting thing about this commutative ring of complex operators
is that it permits us to define for $d=2$ the following exponential mapping%
\begin{equation*}
e^{\left( \alpha \sigma _{0}-i\beta \sigma _{2}\right) }=e^{a}\left(
Cos\beta -i\sigma _{2}Sin\beta \right)
\end{equation*}
where the mathematical structure described in a abstract way before is
clearly seen.

The important thing is that the correct analysis of the algebraic and
divisor ring structure of the BI-field equations is a crucial point to go
towards a truly noncommutative BI theory. The generalization of
transformations (10) is performed due to the fact that the operators can be
realized over the non-commutative field of full-quaternions in the following
manner:%
\begin{equation*}
\frac{1}{\mathbb{R}}\left[ \sigma _{0}\delta ^{\prime }-iG\left( \sigma
_{1}\alpha ^{\prime }+\sigma _{2}\beta ^{\prime }+\sigma _{3}\gamma ^{\prime
}\right) \right]
\end{equation*}%
We assume the coefficients $\alpha ^{\prime },\beta ^{\prime },\gamma
^{\prime }$ :reals and $\delta ^{\prime }$ :complex, in principle, taking
the operator the following form
\begin{equation*}
\frac{1}{\mathbb{R}}\left[ \left(
\begin{array}{cc}
\delta ^{\prime } & 0 \\
0 & ^{\ast }\delta ^{\prime }%
\end{array}%
\right) -iG\left(
\begin{array}{cc}
\gamma ^{\prime } & \alpha ^{\prime }-i\beta ^{\prime } \\
\alpha +i\beta & -\gamma ^{\prime }%
\end{array}%
\right) \right]
\end{equation*}%
where the star means complex conjugation, and the quantity $G$ will have a
different meaning that is in the initial expression (10), obviously. The
question that immediately arises is: Is it possible to impose conditions on
the coefficients $\alpha ^{\prime },\beta ^{\prime },\gamma ^{\prime }$ and $%
\delta ^{\prime }$ in the above expression in order to obtain a
full-quaternionic non-commutative operator from the equations of motion of a
determinant-geometrical action? The answer is affirmative if and only if $%
\gamma ^{\prime }=0$ and $\delta ^{\prime }=\alpha ^{\prime }-i\beta
^{\prime }$. With these particular values of the coefficients the square
root of the determinant in the BI action (where the equations of motion
determining the mapping coming from) is%
\begin{equation*}
\sqrt{-\det (g_{\mu \nu }+b^{-1}\chi F_{\mu \nu })}
\end{equation*}%
where $\chi ^{4}=i(\alpha ^{\prime }-i\beta ^{\prime })$ and $G=\frac{\chi
^{2}}{2b^{2}}F_{\mu \nu }\widetilde{F}^{\mu \nu },$ following the same
conventions from the beginning. A carefully study of the possible physical
meaning will be carried out elsewhere in [11].

\section{Hamiltonian point of view}

We can show that the SO$\left( 2\right) $ structure of the BI theory is more
easily seen in the following operator form [7]:
\begin{equation*}
\frac{1}{R}\left( \sigma _{0}-i\sigma _{2}\overline{\mathbb{P}}\right) L=%
\mathbb{L}
\end{equation*}%
\begin{equation*}
\frac{\mathbb{R}}{\left( 1+\overline{\mathbb{P}}^{2}\right) }\left( \sigma
_{0}+i\sigma _{2}\overline{\mathbb{P}}\right) \mathbb{L}=L
\end{equation*}%
\begin{equation*}
\overline{\mathbb{P}}\equiv \frac{\mathbb{P}}{b}
\end{equation*}%
where we defined the following quaternionic operators:
\begin{equation*}
L=F-i\sigma _{2}\widetilde{F}
\end{equation*}%
\begin{equation*}
\mathbb{L}=\mathbb{F}-i\sigma _{2}\widetilde{\mathbb{F}}
\end{equation*}%
the pseudoescalar of the electromagnetic tensor $F^{\mu \nu }$%
\begin{equation*}
\mathbb{P}=-\frac{1}{4}F_{\mu \nu }\widetilde{F}^{\mu \nu }
\end{equation*}%
where $\sigma _{0}\,$, $\sigma _{2}$ are the well known Pauli matrices that we define
previously. Now, with the definitions given before, we pass to the
description of the phase space from the Hamiltonian point of view in a
similar form as in ref.[5].

The 6-dimensional space $V=\Lambda ^{2}\left( \mathbb{R}^{4}\right)
\rightarrow $2-forms $\in \mathbb{R}^{4}$, has coordinates $F_{\mu \nu }$\
and carries a Lorentz invariant metric with signature $\left( +++---\right) $
defined by%
\begin{equation*}
k_{H}\left( F,F\right) \equiv L\widetilde{L}=2F\widetilde{F}
\end{equation*}%
The dual space $V^{\ast }$ of $V$ consists of skew-symmetric second rank
contravariant tensors $\mathbb{F}^{\mu \nu }$. The phase space $P=V\oplus
V^{\ast }$ carries a natural quaternionic symplectic structure given by%
\begin{equation*}
d\mathbb{L}\wedge d\widetilde{L}=d\mathbb{F}\wedge dF-d\widetilde{F}\wedge d%
\widetilde{\mathbb{F}}
\end{equation*}%
Notice that now, from the mathematical description of the phase space, the $%
SO\left( 2\right) $ symmetry is, in fact, embedded in a large quaternionic
structure.

\section{Maxwell equations in the Riemannian space-time and the Born-Infeld
theory}

We now want to give some curious aspects about the relation between the BI\
field equations and the Maxwell equations in the Riemannian space-time. From
ref.[2] we know that when a gravitational field exists (i.e.:curved
space-time), it is possible to write the Maxwell equations in vacuum
similarly that in a hypotethic medium\footnote{%
Here $g_{\alpha }=-\frac{g_{0\alpha }}{g_{00}},\gamma _{\alpha \beta
}=-g_{\alpha \beta +}\frac{g_{0\alpha }g_{0\beta }}{g_{00}}$ and $h=g_{00}$
as in ref.[2]}

\begin{equation*}
\mathbf{D=}\frac{\mathbf{E}}{\sqrt{h}}\mathbf{+}\left[ \mathbf{B}\times
\mathbf{g}\right] ,\ \mathbf{H=}\frac{\mathbf{B}}{\sqrt{h}}-\left[ \mathbf{E}%
\times \mathbf{g}\right]
\end{equation*}%
(i.e. for a \textit{gyrotropic} medium [3]). Analogously to the Born-Infeld
case, we can put these constitutive relations in the following form\footnote{%
here $\epsilon $ is the full-antisymmetric tensor, as usual}:%
\begin{equation}
\left(
\begin{array}{l}
\mathbf{D} \\
\\
\mathbf{H}%
\end{array}%
\right) ^{\alpha }=\overset{\equiv \mathbb{Q}}{\overbrace{\left[ \frac{%
\sigma _{0}}{\sqrt{h}}+i\sigma _{2}\epsilon _{\beta \gamma }\mathbf{g}%
^{\beta }\right] ^{\alpha }}}\left(
\begin{array}{l}
\mathbf{E} \\
\\
\mathbf{B}%
\end{array}%
\right) ^{\gamma }  \tag{13}
\end{equation}%
Notice a remarkable analogy with the similar expression (10). This means that
the BI
theory can be formulated as an effective metric
theory, as it was shown in references [8]. For the BI case the
constitutive-like relations give $\mathbf{\ D}$ and $\mathbf{H}$ in terms of
$\mathbf{E}$ and $\mathbf{B}$[5]%
\begin{equation*}
\mathbf{D}=\frac{\mathbf{E}+b^{-2}\left( \mathbf{E\cdot B}\right) \mathbf{B}%
}{\sqrt{1+b^{-2}\left( \mathbf{B}^{2}-\mathbf{E}^{2}\right) -b^{-4}\left(
\mathbf{E\cdot B}\right) ^{2}}}
\end{equation*}%
\begin{equation}
\mathbf{H}=\frac{\mathbf{B}-b^{-2}\left( \mathbf{E\cdot B}\right) \mathbf{E}%
}{\sqrt{1+b^{-2}\left( \mathbf{B}^{2}-\mathbf{E}^{2}\right) -b^{-4}\left(
\mathbf{E\cdot B}\right) ^{2}}}  \tag{14}
\end{equation}%
From the introduction we know that these equations can be solved to give E
and H in terms of E and B:%
\begin{equation*}
\mathbf{E}=\frac{\left( 1+b^{-2}\mathbf{B}^{2}\right) \mathbf{D}%
+b^{-2}\left( \mathbf{D\cdot B}\right) \mathbf{B}}{\sqrt{\left( 1+b^{-2}%
\mathbf{B}^{2}\right) \left( 1+b^{-2}\mathbf{D}^{2}\right) -b^{-4}\left(
\mathbf{D\cdot B}\right) ^{2}}}
\end{equation*}%
\begin{equation*}
\mathbf{B}=\frac{\left( 1+b^{-2}\mathbf{D}^{2}\right) \mathbf{B}%
+b^{-2}\left( \mathbf{D\cdot B}\right) \mathbf{D}}{\sqrt{\left( 1+b^{-2}%
\mathbf{B}^{2}\right) \left( 1+b^{-2}\mathbf{D}^{2}\right) -b^{-4}\left(
\mathbf{D\cdot B}\right) ^{2}}}
\end{equation*}%
that make the explicit comparison between (13) and (14) easy when the fields
$\mathbf{D}$ and $\mathbf{H}$ are the same in both the cases: BI fields in
flat space-time and linear field in curved space-time%
\begin{equation*}
\left. \frac{E_{\alpha }}{\mathbb{R}}\right\vert _{BI}=\left. \frac{%
E_{\alpha }}{\sqrt{h}}\right\vert _{f}
\end{equation*}%
\begin{equation*}
\left. \left( \gamma ^{\beta \gamma }B_{\beta }E_{\gamma }\right) \frac{%
B_{\alpha }}{\mathbb{R}}\right\vert _{BI}=\left. \sqrt{\gamma }\epsilon
_{\alpha \beta \gamma }g^{0\alpha }B^{\gamma }\right\vert _{f}
\end{equation*}%
where the subindices $BI$ and $f$ indicate the fields in the BI theory (flat
space-time) and the Maxwell fields in any frame (curved), respectively.

Following the same procedure as in ref.[3] for the Maxwell case without any
background (gravitational and/or electromagnetic), the Fresnel equation in
the Born-Infeld case (in a Lorentz frame) takes the following form:%
\begin{equation*}
-n^{2}\left( C_{xx}n_{x}^{2}+C_{yy}n_{y}^{2}+C_{zz}n_{z}^{2}\right)
+n_{x}^{2}C_{xx}\left( C_{yy}+C_{zz}\right) +
\end{equation*}%
\begin{equation}
+n_{y}^{2}C_{yy}\left( C_{xx}+C_{zz}\right) +n_{z}^{2}C_{zz}\left(
C_{xx}+C_{yy}\right) -C_{xx}C_{yy}C_{zz}=0  \tag{15}
\end{equation}%
where $n_{i}$ are the coordinates of the surface of propagation (wave
number) and
\begin{equation}
C_{ij}\equiv \frac{\left( \delta _{ij}+\left( \mathbf{E\cdot B}\right)
E_{i}B_{j}\right) }{1+b^{-2}\left( \mathbf{B}^{2}-\mathbf{E}^{2}\right)
-b^{-4}\left( \mathbf{E\cdot B}\right) ^{2}}  \tag{16}
\end{equation}%
Notice that expression (15)\ has the same form as in reference [3] but with $%
\varepsilon _{ij}$ replaced by $C_{ij}$ given by (16). Notice also that in
the presence of any electromagnetic background the particular form of the
Fresnel equation (15) can take a more general form depending on the
components for $C_{ij}$ with $c\neq j$ (i.e. ref.[3]). It is also
interesting as a theoretical tool to test the nonlinearity of the BI field
as the deviation of the Maxwell theory. This fact takes particular
importance in astrophysical phenomena [10].

\section{Concluding remarks}

In this work, the Born-Infeld field equations were written in the most
compact form by means of quaternionic operators constructed according to the
symmetries of the theory.

We also showed that the $\mathbb{Q}$ operators defined here form part of a
commutative ring of complex operators and the $SO\left( 2\right) $ symmetry
of the BI field equations is in a manner embedded into a larger quaternionic
structure. This extension can be realized by transforming the commutative
ring of complex operators to a non-commutative ring. Our results agree with
the observation of Gibbons and Rasheed in [5,6] that there exists discrete
symmetry in the structure of the field equations that is unique in the case
of the Born and Infeld nonlinear electrodynamics: this fact is easily seen
in our work because these discrete symmetries generated by the $\mathbb{Q}$
operators are invertible.

The quaternionic structure of the phase space was explicitly derived and
described from the Hamiltonian point of view, showing at the same time that
the results on the structure of the phase space of ref.[5] are naturally
included in this large quaternionic symmetry.

Finally, the analogy between the BI\ theory and the Maxwell (linear)
electrodynamics in a curved space-time was explicitly shown and the equation
for the Fresnel equation in the nonlinear BI case without background was
explicitly given and proposed as a theoretical tool to test this
particularly interesting nonlinear electrodynamics of M. Born and L. Infeld.

\section{Acknowledgements}

I am very thankful to Professors Boris M. Barbashov and Alexander Dorokhov
for their advisement and, in particular, to Professor J. A. Helayel-Neto for
his interest to put in this research. I am very grateful to the
JINR-Directorate, in particular, the Bogoliubov Laboratory of Theoretical
Physics for their hospitality and support.

\section{References}

[1] M. Born and L. Infeld, Proc. Roy. Soc.(London) \textbf{144}, 425 (1934).

[2] L.D. Landau and E.M. Lifshitz, \textit{Teor\'{\i}a cl\'{a}sica de los
Campos}, (Revert\'{e}, Buenos Aires, 1974).

[3] L.D. Landau and E.M. Lifshitz, \textit{Electrodinamica de los medios
continuos}, (Revert\'{e}, Buenos Aires, 1974).

[4]. M. K. Gaillard and B. Zumino, Nucl. Phys. \textbf{B 193}, 221 (1981).

[5] G. W. Gibbons and D. A. Rasheed, Nucl. Phys. \textbf{B 454}, 185 (1995).

[6] G. W. Gibbons and D. A. Rasheed, Phys. Lett. \textbf{B 365}, 46 (1996).

[7]. D. J. Cirilo-Lombardo, Class. Quntum. Grav. 21, 1407 (2004), and
references therein.

[8].M. Novello et Al., Phys. Rev. \textbf{D 61}, 45001 (2000), and
references therein.

[9].F. P. Schuller, Annals of Phys.\textbf{\ 299}, 174 (2002).

[10].Y. B. Zeldovich, JETP Lett. \textbf{1}, 95 (1965).

[11]. D. J. Cirilo-Lombardo, work in preparation.

\end{document}